\documentclass[twocolumn,prd,nofootinbib,superscriptaddress,eqsecnum,tightenlines,8pt]{revtex4}

\usepackage{fancyhdr}
\usepackage{amsmath,amsfonts,amssymb}
\usepackage[colorlinks,citecolor=blue,linkcolor=blue,urlcolor=blue]{hyperref}
\usepackage[english]{babel}

\usepackage{color}
\definecolor{red}{rgb}{1,0,0}
\definecolor{gre}{rgb}{0,0.6,0}
\definecolor{blu}{rgb}{0,0,1}

\usepackage{mathenv}
\usepackage{ulem}

\usepackage[pdftex]{graphicx}
\usepackage{mathrsfs} 
\def\be{\begin{equation}}
\def\ee{\end{equation}}

\begin{document}

\title{Detailed investigation of the duration of inflation in loop quantum cosmology for a Bianchi I universe with different inflaton potentials and initial conditions.}

\date{\today}

\author{Killian Martineau}
\email{martineau@lpsc.in2p3.fr}
\affiliation{
Laboratoire de Physique Subatomique et de Cosmologie, Universit\'e Grenoble-Alpes, CNRS-IN2P3\\
53, Avenue des Martyrs, 38026 Grenoble cedex, France\\
}%

\author{Aur\'elien Barrau}
\email{Aurelien.Barrau@cern.ch}
\affiliation{
Laboratoire de Physique Subatomique et de Cosmologie, Universit\'e Grenoble-Alpes, CNRS-IN2P3\\
53, Avenue des Martyrs, 38026 Grenoble cedex, France\\
}%

\author{Susanne Schander}
\email{susanne.schander@gravity.fau.de}
\affiliation{Institute for Quantum Gravity, University of Erlangen-N\"{u}rnberg \\
Staudtstr. 7B, D-91058 Erlangen, Germany
}%

\begin{abstract}
There is a wide consensus on the correct dynamics of the background in loop quantum cosmology. In this article we make a systematic investigation of the duration of inflation by varying what we think to be the most important ``unknowns" of the model: the way to set initial conditions, the amount of shear at the bounce and the shape of the inflaton potential.
\end{abstract}

\maketitle

\section{Introduction}

Loop quantum gravity (LQG) is a promising attempt to perform a nonperturbative background-invariant quantization of general relativity (GR). General reviews can be found, {\it e.g.}, in \cite{lqg1,lqg2,lqg3,lqg4,lqg5,lqg6,lqg7,lqg8,lqg9,lqg10}. Loop quantum cosmology (LQC) is a quantum theory inspired by LQG that takes into account the cosmological symmetries. Some recent reviews can be found, {\it e.g.}, in \cite{lqc1,lqc2,lqc3,lqc4,lqc5,lqc6,lqc7,lqc8,lqc9,lqc10,lqc11,lqc12}. The status of perturbations in LQC is still not fully clear. On the one hand, the {\it deformed algebra} approach, which puts the emphasis on the consistency of the effective gauge theory, has been investigated in detail (see, {\it e.g.}, \cite{Bojowald:2011aa,eucl3,tom1,tom2,eucl2,lcbg,Schander:2015eja}). On the other hand, the {\it dressed metric} approach, which puts the emphasis on the quantum treatment of the background and the perturbations, has been pushed forward (see, {\it e.g.}, \cite{Agullo1,Agullo2,Agullo3}). Other attempts have also been suggested, for example in \cite{ed} and \cite{merce}. At this stage, there is no wide consensus on LQC predictions for the primordial power spectra although some general trends can be underlined \cite{Bolliet:2015bka}.\\

Concerning the dynamics of the LQC background however, different approaches lead to the very same dynamical equations, underlining the robustness of the model. The effective modified Friedmann equation, 
\begin{equation}
H^2=\frac{\kappa}{3}\rho\left(1-\frac{\rho}{\rho_c}\right),
\end{equation}

is one of the general predictions of LQC.
In this equation $H$ stands for the Hubble parameter, $\rho$ for the energy density, $\rho_c\sim \rho_{Pl}$ for the maximum energy density, and $\kappa= 8 \pi$.  Beyond the standard Hamiltonian LQC calculation, the above equation has even been rederived in quantum reduced loop gravity \cite{Alesci:2014rra} and in group field theory \cite{Gielen:2013kla,Gielen:2013naa} (with a possible slight shift in the bounce energy). In this article, we focus on this robust background dynamics. Remarkably, in this cosmological paradigm, inflation occurs naturally, this being a consequence of the strong attractor status of its solutions when one considers a scalar field as the content of the Universe. Probably, the most interesting output of the LQC framework is that the duration of inflation itself can, to some extent, be predicted. \\

Still, even at the background level, three main uncertainties remain to be addressed systematically. The first one is the way to choose initial conditions. There are two schools of thought: one sets them in the remote past of the contracting branch, and the other one sets them at the bounce. The important question here is not related with the conditions themselves (they are in a one-to-one correspondence with one another), but with the variable to which a known (and presumably flat) probability distribution function (PDF) can be assigned. This is an important conceptual issue that will be discussed later in this article. The second uncertainty is associated with the amount of anisotropic shear at the bounce. As it will be diluted very fast during the expansion it might be very high at the bounce and remain compatible with observational data. In this study, we focus on the Bianchi I dynamics and consider different contributions from the shear. Since anisotropies scale as $a^{-6}$ in a Bianchi I universe, $a$ being the scale factor, they inevitably grow during the contracting phase and they are expected to play an important role in any bouncing model.  The third main uncertainty is associated with the inflaton potential as LQG does not make any predictions concerning the matter content of the Universe. So far, the status is unclear and the matter content has to be assumed independently. In this paper we focus on four different potentials which are favored by the latest Planck results \cite{Planck2015}.

\section{Formalism}

The metric for a homogeneous Bianchi I universe is given by

\begin{equation}
ds^{2} = -dt^{2}+a_{1}^{2}dx^{2} +a_{2}^{2}dy^{2}+a_{3}^{2}dz^{2}.
\end{equation}

Anisotropies appear through three independent directional scale factors $\left\lbrace a_1, a_2, a_3 \right\rbrace$.

The spatial hypersurface $\Sigma$ of this spacetime has an $\mathbb{R}^{3}$ topology. Since it is not compact, many spatial integrals will diverge, but one can use the fundamental property of homogeneous spaces to restrict the study to a fiducial cell $\mathcal{V}$ on the spatial manifold which will not appear in the final results. Its finite fiducial volume is given by $V_{0} = l_{1} l_{2} l_{3}$, and its edges are chosen to lie along the fiducial orthonormal triads $\overset{\circ}{e^{a}_{i}}$. Fiducial orthonormal cotriads $\overset{\circ}{\omega^{i}_{a}}$ are also introduced in a such a way that the fiducial spatial metric can be written as $\overset{\circ}{q}_{ab} = \overset{\circ}{\omega^{i}_{a}} \overset{\circ}{\omega^{j}_{b}} \delta_{ij}$. The Ashtekar connection $A^{i}_{a}$ and the densitized triads $E_{i}^{a}$ can be reduced using the symmetries of the spatial manifold of the homogeneous Bianchi I spacetime:

\begin{equation}
A^{i}_{a} = c^{i} (l_{i})^{-1} \overset{\circ}{\omega^{i}_{a}} ~~~ \text{and} ~~~ E_{i}^{a} = \dfrac{p_{i} l_{i}}{V_{0}} \sqrt{det \left(\overset{\circ}{q}_{ab}\right)} \overset{\circ}{e^{a}_{i}},
\end{equation}

where the coefficients $c^{i}$ and $p_{i}$ are the symmetry-reduced coefficients of the Ashtekar connection and of the densitized triad. They form a canonical set with the following Poisson brackets:

\begin{equation}
\left\lbrace c^{i} , p_{j} \right\rbrace = \kappa \gamma \delta^{i}_{j},
\end{equation}
where $\gamma= 0.2375$ is the Barbero-Immirzi parameter whose value has been obtained by evaluating the black hole entropy in LQG \citep{BlackHoleEntropy}. The specific choice of this parameter is still a source of debates, but the precise numerical value is not fundamental for the study presented here (the $\gamma$-dependence of the energy density available at the bounce is quite trivial).

The $p_{i}$ coefficients can be expressed in terms of the cosmological directional scale factors:

\begin{equation}
\left\{
    \begin{aligned}
p_{1} = \epsilon_{1} l_{2} l_{3} |a_{2}a_{3}|, \\
p_{2} = \epsilon_{2} l_{1} l_{3} |a_{1}a_{3}|, \\
p_{3} = \epsilon_{3} l_{1} l_{2} |a_{1}a_{2}|, 
\end{aligned}
  \right.
\end{equation}

where $\epsilon_{i} = \pm 1$ depending on the orientation of the triads. Without any loss of generality, we fix $\epsilon_{i} = +1$ and $l_{i}= 1$, leading to $V_{0}=1$.

The directional scale factors can  be written in terms of the reduced densitized triads:

\begin{equation}
a_{1}=\sqrt{\dfrac{p_{2}p_{3}}{p_{1}}}, ~~~~~ \text{and cyclic expressions,}
\end{equation}

leading to the directional Hubble parameters,

\begin{multline}
H_{1}:=\dfrac{\dot{a}_{1}}{a_{1}}=-\dfrac{\dot{p}_{1}}{2p_{1}}+\dfrac{\dot{p}_{2}}{2p_{2}}+\dfrac{\dot{p}_{3}}{2p_{3}}, \\ \text{and cyclic expressions,}
\end{multline}

where the dots refer to derivatives with respect to cosmic time.

We define a mean scale factor,

\begin{equation}
a:=(a_{1}a_{2}a_{3})^{1/3},
\end{equation}

in order to obtain a mean Hubble parameter

\begin{equation}
H:=\dfrac{\dot{a}}{a}=\dfrac{1}{3}(H_{1}+H_{2}+H_{3}).
\end{equation}

The classical evolution of the metric is given by the following Hamiltonian:

\begin{equation}
\mathcal{H} = \mathcal{H}_{Grav}(c_{i},p_{i}) + \mathcal{H}_{M}(p_{i}, \Phi, \pi),
\end{equation} 

where $\Phi$ is a scalar field, and $\pi$ is its conjugate momentum. The gravitational and matter Hamiltonians are respectively given by \citep{AshtekarHamiltonians}

\begin{equation}
\mathcal{H}_{Grav} = - \dfrac{N}{\kappa \gamma^{2}} \left(a_{1}c_{2}c_{3}+a_{2}c_{1}c_{3}+a_{3}c_{1}c_{2} \right),
\end{equation} 

and

\begin{equation}
\mathcal{H}_{M}=N\sqrt{p_{1}p_{2}p_{3}}\rho,
\end{equation}

where $N$ is the lapse function.\\

Quantization of the above cosmological model within the lines of LQC requires the introduction of \textit{holonomy corrections}. At the effective level, this procedure basically consists of the following replacement:

\begin{equation}
c_{i} \rightarrow \dfrac{\sin(\bar{\mu}_{i}c_{i})}{\bar{\mu}_{i}},
\end{equation}

where $\bar{\mu}_{i}$ are given by

\begin{equation}
\bar{\mu}_{i} = \dfrac{\lambda}{a_{i}},
\end{equation}

with $\lambda = \sqrt{\Delta} = \sqrt{4\sqrt{3}\pi \gamma}$, the square root of the minimum eigenvalue of the area operator in LQG. 

We introduce three fundamental parameters $h_{i}$:

\begin{equation}
h_i := \bar{\mu}_{i}c_{i}=\dfrac{\lambda c_{i}}{a_{i}}.
\end{equation}

Those three parameters are gauge-invariant variables which can be interpreted as the classical limits of the quantum equivalents of the directional Hubble parameters.
After implementing the holonomy corrections, the effective gravitational Hamiltonian becomes

\begin{eqnarray}
\mathcal{H}_{Grav} & = & - \dfrac{N \sqrt{p_{1}p_{2}p_{3}}}{\kappa \gamma^{2} \lambda^{2}} [\sin(h_{1})\sin(h_{2}) \\ \nonumber
& & + \sin(h_{2})\sin(h_{3}) + \sin(h_{1})\sin(h_{3}) ].
\end{eqnarray}

Besides, the functional form of the matter Hamiltonian does not get changed as the matter Hamiltonian does not depend on the $c_{i}$ coefficients. We therefore assume that it remains unchanged by the quantization procedure.\\

Following the pioneering work of \cite{Gupt:2013swa} and rewriting the effectively quantized Hamiltonian constraint, $\mathcal{H}=0$, one can find the generalized Friedmann equation for a Bianchi I universe with holonomy corrections \citep{LindaFriedmann}:

\begin{equation}
H^{2} = \sigma_{Q}^{2}+\dfrac{\kappa}{3}\rho-\lambda^{2}\gamma^{2}\left(\dfrac{3}{2}\sigma_{Q}^{2}+\dfrac{\kappa}{3}\rho\right)^{2},
\label{FriedmannLQC}
\end{equation}

where $\sigma_{Q}^{2}$ corresponds to the quantum shear and can be expressed in terms of the $h_{i}$ coefficients:

\begin{eqnarray}
\label{SigmaQ(hi)}
\sigma_{Q}^{2} & := & \dfrac{1}{3\lambda^{2}\gamma^{2}}\Bigg( \Bigg. 1-\dfrac{1}{3}\big[ \big. cos(h_{1}-h_{2}) \\ \nonumber 
 & & +cos(h_{2}-h_{3})+cos(h_{3}-h_{1}) \big. \big]\Bigg. \Bigg).
\end{eqnarray}

It should be stressed that the way anisotropies are defined here, in agreement with \cite{LindaFriedmann}, differs from the usual cosmological definition.
Upper limits for $\rho$ and $\sigma_{Q}^{2}$ can easily be obtained by requiring $H^{2}>0$ in Eq. (\ref{FriedmannLQC}):

\begin{equation}
\rho \leqslant \rho_{c} = \dfrac{3}{\kappa \lambda^{2} \gamma^{2}}, ~~~~~ \text{obtained when } \sigma_{Q}^{2}=0 ~ ,
\end{equation}

\begin{equation}
\sigma_{Q}^{2} \leqslant \sigma_{Q_{c}}^{2} = \dfrac{4}{9\lambda^{2}\gamma^{2}}, ~~~~~ \text{obtained when } \rho=0 ~ .
\end{equation}

The dynamics of the $p_{i}$-functions is given by 

\begin{multline}
\dot{p}_{1} = \dfrac{1}{N} \left\lbrace p_{1}, \mathcal{H} \right\rbrace = \dfrac{p_{1}}{\gamma \lambda } \cos(h_{1}) \left[ \sin(h_{2}) + \sin(h_{3}) \right], \\ \text{and cyclic expressions.}
\end{multline}

From this, the classical directional Hubble parameters, $H_{i}$, can be expressed as functions of the $h_{i}$'s:

\begin{multline}
H_{1} = -\dfrac{\dot{p}_{1}}{2p_{1}}+\dfrac{\dot{p}_{2}}{2p_{2}}+\dfrac{\dot{p}_{3}}{2p_{3}} \\ = \dfrac{1}{2 \gamma \lambda} \left[\sin(h_{1}-h_{2}) + \sin(h_{1}-h_{3}) + \sin(h_{2}+h_{3}) \right],\\ \text{and cyclic expressions.}
\label{Hi(hi)}
\end{multline}

The total Hubble parameter then reads

\begin{multline}
H = \dfrac{1}{6 \gamma \lambda} \left[\sin(h_{1}+h_{2}) + \sin(h_{1}+h_{3}) + \sin(h_{2}+h_{3}) \right].
\label{H(hi)}
\end{multline}

In the same way, the dynamics of the $h_{i}$'s is given by the following equations:

\begin{eqnarray}
\label{DynhiLQC}
\dot{h}_{1} & = & \dfrac{1}{N} \left\lbrace h_{1},\mathcal{H} \right\rbrace \\ \nonumber 
& = & \dfrac{1}{2\gamma\lambda} \big[ \big. (h_{2}-h_{1})(\sin(h_{1})+\sin(h_{3}))\cos(h_{2}) \\ \nonumber
& & + (h_{3}-h_{1})(\sin(h_{1})+\sin(h_{2}))\cos(h_{3}) \big. \big] 
\\ \nonumber
& & - \dfrac{\kappa\gamma\lambda}{2} (\rho + P) ~~~ \text{and cyclic expressions,}
\end{eqnarray}

where the pressure $P$ is defined to fulfill the continuity equation $\dot{\rho} = 3 H ( \rho + P ) $.

In this study, the matter content of the Universe is assumed to be a scalar field $\Phi(t)$. Its evolution is given by the Klein-Gordon equation:

\begin{equation}
\ddot{\Phi} + 3 H \dot{\Phi} + \dfrac{dV}{d\Phi} = 0.
\label{KleinGordon}
\end{equation}

The previous equations drive the dynamics of the system. They are the basis for the subsequent simulations.

\section{Simulations}

\subsection{Description of the chosen potentials}

For the purpose of this study, we choose four different potentials, which are all in good agreement with the most recent Planck data \citep{Planck2015} as far as standard cosmological models are concerned.

\begin{itemize}

\item The most common potential when dealing with slow-roll inflation is the quadratic one: 
\begin{equation}
V(\Phi)=\dfrac{1}{2}m^{2}\Phi^{2} .
\end{equation}

Although it is not the best fit to the most recent CMB measurements, a massive scalar field is very useful in order to compare  different approaches. For this potential, we fix $m_{\text{quadratic}}=1.21 \times 10^{-6} m_{\text{Pl}}$, as suggested by the Planck data \citep{Planck2015}.

\item The large tensor-to-scalar ratio $r$ initially reported by BICEP2 \citep{BICEP2} can be generated by an inflation based on a simple monomial effective potential $V(\Phi) \propto \Phi^{p}$. Although the initial analysis was shown to be incorrect and values $p>2$ are now strongly disfavored by Planck \citep{Planck2015}, some values of $p<2$, like $p= 2/3$, $p=1$ or $p=4/3$ are still in good agreement with the data. In addition to the quadratic potential previously mentioned, we  therefore explore the LQC dynamics with the potential associated with $p=1$:

\begin{equation}
V(\Phi)= \Lambda^{3} \Upsilon \vert \Phi \vert,
\end{equation} 

with the following parametrization: $\Lambda = 1.23 \times 10^{-3}$ and $\Upsilon = 1.22 \times 10^{-1}$ \citep{PowerLawPotential}.
The mass of the scalar field with this potential is given by $ m_{\text{monomial,p=1}} \sim \Lambda \times \Upsilon \sim 1.50 \times 10^{-4} m_{\text{Pl}}$ \citep{PowerLawPotential}.

\item Inflation can also be motivated by supergravity and string theory. In the context of type IIB string compactifications, and with a simple string model of inflation, the effective inflaton potential is well approximated by \citep{StringPotential}:  

\begin{eqnarray}
V(\Phi) & \simeq & \dfrac{C_{2}}{\left\langle \nu \right\rangle ^{10/3} } \Bigg[ \Bigg. (3 - R) - 4 (1 + \dfrac{1}{6} R ) e^{-\dfrac{\Phi}{\sqrt{3}}} \\ \nonumber
& + & (1 + \dfrac{2}{3} R) e^{-\dfrac{4 \Phi}{\sqrt{3}}} + R e^{\dfrac{2 \Phi}{\sqrt{3}}} \Bigg. \Bigg],
\end{eqnarray}

where the following parametrization has been chosen: $C_{2} = 5157.35$, $R=2.3 \times 10^{-6}$ and $\left\langle \nu \right\rangle = 1709.55$ \citep{StringPotential}.
The mass $m_{\text{stringy}}=5.87 \times 10^{-4} m_{\text{Pl}}$ of the inflaton field is given by the curvature of the potential around its minimum $V''(0)$. Although this study is focused on LQG, we investigate this string-inspired potential as a good phenomenological description of inflation.

\item The last potential we will focus on is the Starobinsky potential. Even if the statistical significance of this statement is to be taken with care, models with the Starobinsky potential have the best accordance \cite{Martin:2013tda} with observational data \citep{Planck2015}. The potential is given by
 
\begin{equation}
V(\Phi)=\dfrac{3 m^{2}}{4 \kappa} \left(1- e^{-\sqrt{\dfrac{2 \kappa}{3}}\Phi} \right)^{2}.
\end{equation}

The mass value for this potential is fixed to be $m_{\text{Starobinsky}} = 2.51 \times 10^{-6} m_{\text{Pl}} $ \citep{Bonga2016}. \\

\end{itemize}

\begin{figure}[!h]
\begin{center}
\includegraphics[scale=0.60]{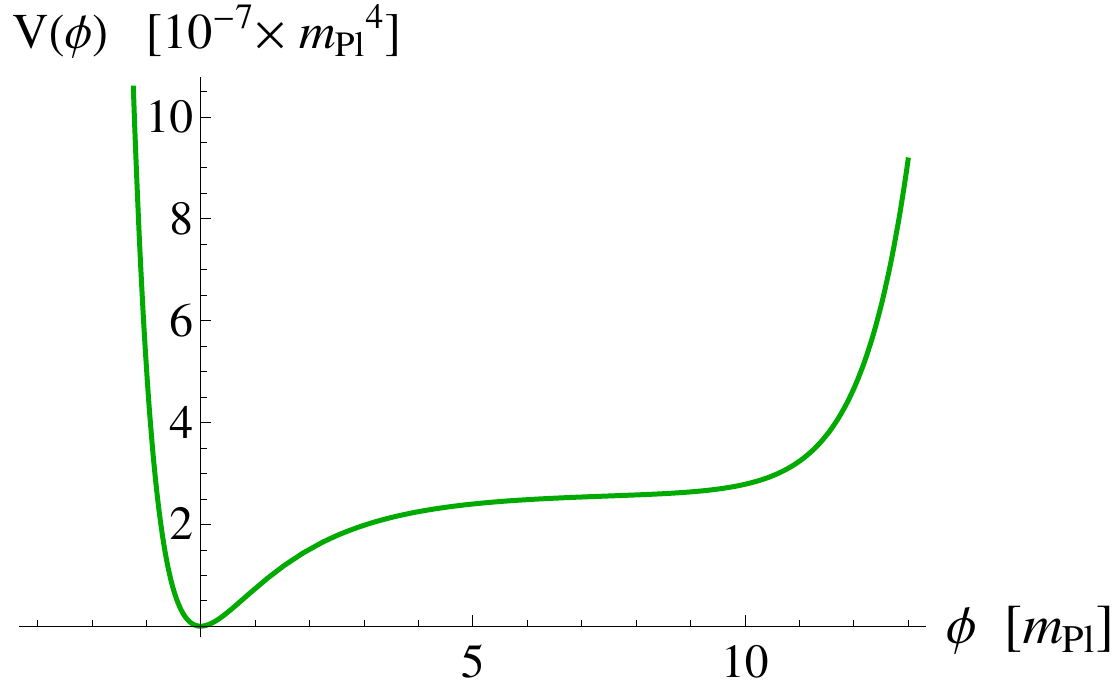}
\includegraphics[scale=0.60]{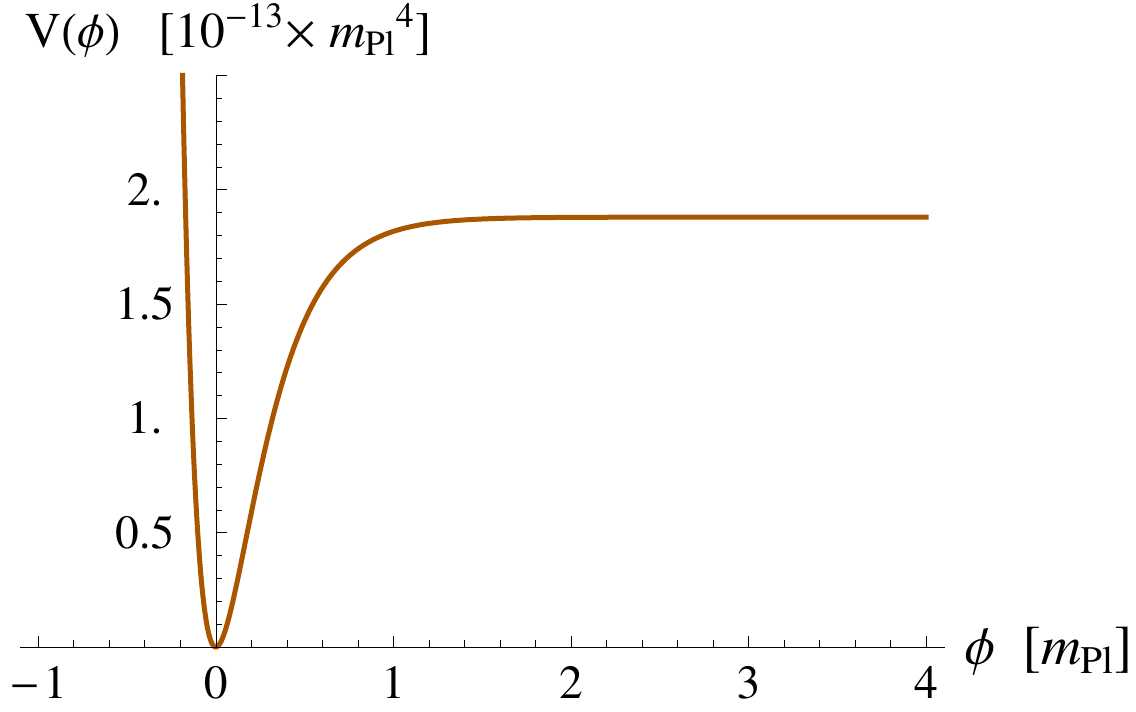}
\caption {\underline{Upper panel}: String-theory-inspired inflaton potential according to the chosen parametrisation. \underline{Lower panel}: Starobinsky potential for a mass of the inflaton field $m_{\text{Starobinsky}} = 2.51 \times 10^{-6} m_{\text{Pl}} $. } 
\label{Plots Potentiels}
\end{center}
\end{figure}

The shapes of the string-inspired potential and of the Starobinsky potential are displayed in Fig. \ref{Plots Potentiels}.

\subsection{Duration of slow-roll inflation} 

Once the inflaton potential $V(\Phi)$ has been chosen, the key question to address is the one of the associated duration of inflation for the given initial conditions.

For this purpose, we express the number of \textit{e}-folds of slow-roll inflation as the integral

\begin{equation}
N = \int_{a_{i}}^{a_{f}} d\ln(a) = \vert \int_{\Phi_{i}}^{\Phi_{f}} \dfrac{1}{\sqrt{2 \epsilon_{V}}} \sqrt{\kappa} \dfrac{ d\Phi}{m_{Pl}} \vert ~.
\end{equation}

In this expression, $\Phi_{i}$ stands for the value of the scalar field at the beginning of the slow-roll phase and $\Phi_{f}$ is such that $\epsilon_{V}(\Phi_{f}) = 1$, where 

\begin{equation}
\epsilon_{V}(\Phi) \equiv \dfrac{1}{2 \kappa} \left( \dfrac{V_{,\Phi}}{V} \right)^{2} m_{pl}^{2}  ~
\end{equation}

is the first slow-roll parameter which is equivalent to the first Hubble flux parameter under slow-roll assumptions.

This expression for $N$ leads to the following results for the different potentials considered in this study:

\begin{equation}
\textit{Quadratic potential} :  N = 2 \pi \Phi_{i}^{2} - \dfrac{1}{2} ~ ,
\end{equation}

\begin{equation}
\textit{Linear potential} :  N  = 4 \pi \Phi_{i}^{2} -\dfrac{1}{4} ~ ,
\end{equation}

\begin{eqnarray}
&& \textit{Starobinsky potential} :  \\ \nonumber  N & = & \dfrac{3}{4} \ln\left(1+\dfrac{2}{\sqrt{3}} \right) - \dfrac{3}{4} \left(1+\dfrac{2}{\sqrt{3}} \right) \\ \nonumber & - & \sqrt{\dfrac{3 \kappa}{8}} \Phi_{i} + \dfrac{3}{4} e^{\sqrt{\dfrac{2 \kappa}{3}} \Phi_{i}} ~ .
\end{eqnarray}

In the case of the string theory potential the integral is computed numerically. \\

The last ingredient needed to fully describe the dynamics of the Universe is the choice of a set of initial conditions. As mentioned in the Introduction, there are two main schools of thought about the way to implement initial conditions in LQC. The first line of thought \cite{bl,LindaInflation} follows the argumentation that setting initial conditions in the remote past makes sense since it is the classical phase where physics is well under control, and this is logically consistent if causality is to be taken seriously. In addition, there is then a variable to which a flat PDF can naturally be assigned: the phase of the oscillations of the scalar field. This flat PDF is, in addition, preserved over time when quantum corrections remain small. The other point of view \cite{AS2011} is to set initial conditions at the bounce, which is the only special moment in the cosmic history. The relevant variable to which one can assign a flat PDF is then the fraction of potential energy at the bounce. In the following, we will study both possibilities and investigate the effects of anisotropies in each case. We will, however, argue that setting initial conditions in the remote past is in our opinion more consistent.

\subsection{Initial conditions in the remote past}

Using a Taylor expansion, we assume that all potentials can be approximated by a quadratic form far enough from the bounce in the classical contracting phase. This is possible because when the energy density is very small, as expected in the remote past of the prebounce branch, the field is near the bottom of its potential.

\subsubsection{Initial conditions for the matter sector}

In order to describe the evolution of the scalar field, we introduce two dynamical parameters, the potential energy parameter $x$ and the kinetic energy parameter $y$, defined by

\begin{equation}
x(t) := \sqrt{\dfrac{V(\Phi)}{\rho_{c}}} ~, ~~~~~ y(t) := \sqrt{\dfrac{\dot{\Phi}^{2}}{2\rho_{c}}}~.
\end{equation}

They satisfy

\begin{equation}
x^{2}(t)+y^{2}(t)=\dfrac{\rho(t)}{\rho_{c}} ~.
\end{equation}

In the case of the quadratic potential, $x(t)$ becomes

\begin{equation}
x(t) = \dfrac{m \Phi(t)}{\sqrt{2 \rho_{c}}} ~.
\end{equation}

The Klein-Gordon equation (\ref{KleinGordon}) can therefore be written as

\begin{equation*}
  \left\{
    \begin{aligned}
&  \dot{x} = m y, \\
& \dot{y} = -3Hy-mx.
    \end{aligned}
  \right.
  \label{Klein Gordon 2 eqs}
\end{equation*}

The evolution of the scalar field is driven by two different time scales; the classical one $1/m$, and the quantum one $1/\sqrt{3 \kappa \rho_{c}}$. The ratio of these two time scales is given by

\begin{equation}
\Gamma := \dfrac{m}{\sqrt{3 \kappa \rho_{c}}} .
\end{equation}

In the classical phase before the bounce, we assume that the following conditions are satisfied:

\begin{equation}
H(t) <0 ~~, ~~~ \sigma_{Q}^{2}(t) \ll \dfrac{\kappa}{3}\rho(t) ~~~ \text{and} ~~~ \sqrt{\dfrac{\rho(t)}{\rho_{c}}} \ll \Gamma ~.
\label{Hypotheses phase classique}
\end{equation}

As long as the assumption $\sqrt{\rho / \rho_{c} } \ll \Gamma$ holds, the Klein-Gordon equation (\ref{KleinGordon}) reduces to the one of a simple harmonic oscillator, and $x$ and $y$ are thus given by

\begin{equation}
  \left\{
    \begin{aligned}
  x(t) \simeq \sqrt{\dfrac{\rho(0)}{\rho_{c}}} \sin(mt+\delta), \\
  y(t) \simeq \sqrt{\dfrac{\rho(0)}{\rho_{c}}} \cos(mt+\delta).
    \end{aligned}
  \right.
  \label{x(t) y(t) oscillants}
\end{equation}

The $\delta$-parameter, i.e. the phase of the oscillating scalar field, plays an important role in this study.
Still under the hypothesis given by Eq. (\ref{Hypotheses phase classique}), and by using the derivative of the Friedmann equation restricted to lowest order terms in $x$ and $y$, one obtains the expression for the energy density:

\begin{equation}
\rho(t) \simeq \rho_{c} \left(\dfrac{\Gamma}{\alpha}\right)^{2} \left[ 1 - \dfrac{1}{2 \alpha}  \left( mt + \dfrac{1}{2} \sin( 2 m t + 2 \delta) \right) \right]^{-2},
\label{rho(t)}
\end{equation}

where $\alpha$ is a free parameter set to ensure that Eq. (\ref{Hypotheses phase classique}) remains valid. It has been shown in \citep{LindaValeurAlpha} that the shape of the PDF of the duration of slow-roll inflation does not depend on the value of $\alpha$ as long as it is high enough. For the purpose of this study, we have chosen $\alpha = 17/4 \pi +1$. This value induces enough oscillations of the field in the contracting phase (more than 10) and is convenient to derive analytical solutions in the case of the quadratic potential.

Setting $t=0$ in Eqs. (\ref{x(t) y(t) oscillants}) and (\ref{rho(t)}) gives the initial conditions for the matter sector:

\begin{equation}
  \left\{
    \begin{aligned}
&  \Phi (0) = \sqrt{2 \rho(0)} \sin(\delta )/m, \\
&  \dot{\Phi} (0) = \sqrt{2 \rho(0)} \cos(\delta ),
    \end{aligned}
  \right.
\end{equation}

and

\begin{equation}
\rho(0) = \rho_{c} \left(\dfrac{\Gamma}{\alpha}\right)^{2} \left[ 1 - \dfrac{1}{4 \alpha} \sin(2 \delta) \right]^{-2} .
\end{equation}

Since we have no constraint on the initial PDF of the quantum shear $\sigma_{Q}^{2}(0)$, except that it must fulfill Eq. (\ref{Hypotheses phase classique}), we express the initial quantum shear as a fraction of the initial energy density:

\begin{equation}
\sigma_{Q}^{2}(0)= f \dfrac{\kappa}{3} \rho(0) .
\end{equation}

The parameter $f \ll 1$ represents the ratio of the initial quantum shear over the initial energy density.\\

For fixed values of $\alpha$ and f, the only free variable which remains to be chosen in order to fix the initial parameters $\lbrace \Phi(0),\dot{\Phi}(0),\rho(0),\sigma_{Q}^{2}(0)\rbrace$ completely, is the initial phase of the scalar field $\delta$. The question of how to fix $\delta$ is therefore crucial to determine the dynamics. The most reasonable PDF choice for the $\delta$-parameter is a flat one, since the phase of the field is purely contingent without any physically preferred value. Most importantly, as shown in \citep{LindaValeurAlpha}, and as explained before, this PDF is preserved over time as long as one does not approach the bouncing phase. The fact that there exists a specific variable to which a physically well-motivated PDF can be assigned is a very important feature of the model. This is the main reason why predictions for the duration of inflation can be made.

\subsubsection{Initial conditions for the background dynamics}

Far before the bounce, one can approximate Eqs. (\ref{SigmaQ(hi)}) and (\ref{H(hi)}) by their Taylor development at first order. This leads to the following initial conditions:

\begin{equation}
H(0) \simeq \dfrac{1}{3 \gamma \lambda} \left( h_{1}(0) + h_{2}(0) + h_{3}(0) \right),
\label{H(0)}
\end{equation}

and   

\begin{eqnarray}
\label{Sigma(0)(hi)}
\sigma_{Q}^{2}(0) & \simeq & \dfrac{1}{18 \gamma^{2} \lambda^{2}} \big[ \big. \left(h_{1}(0) - h_{2}(0) \right)^{2} \\ \nonumber
& & + \left(h_{1}(0) - h_{3}(0) \right)^{2} + \left(h_{2}(0) - h_{3}(0) \right)^{2} \big. \big].
\end{eqnarray}

We define a symmetry variable for the anisotropy:

\begin{equation}
S := \dfrac{(h_{2}-h_{1})-(h_{3}-h_{2})}{(h_{3}-h_{1})}.
\label{Symmetry parameter}
\end{equation}

Without any loss of generality, we choose the following labeling

\begin{equation}
h_{1} \leq h_{2} \leq h_{3} ,
\end{equation}

such that $ 0 \leq \vert S \vert \leq 1$. 

Solving Eqs. (\ref{H(0)}) and (\ref{Sigma(0)(hi)}) with Eq. (\ref{Symmetry parameter}) provides the initial conditions for the $h_{i}$-parameters:

\begin{equation}
  \left\{
    \begin{aligned}
  h_{1}(0) \simeq \gamma \lambda H(0) - \gamma \lambda \dfrac{3+S}{\sqrt{3+S^{2}}}\sqrt{\sigma_{Q}^{2}(0)}, \\
  h_{2}(0) \simeq \gamma \lambda H(0) + \gamma \lambda \dfrac{2S}{\sqrt{3+S^{2}}}\sqrt{\sigma_{Q}^{2}(0)}, \\
  h_{3}(0) \simeq \gamma \lambda H(0) + \gamma \lambda \dfrac{3-S}{\sqrt{3+S^{2}}}\sqrt{\sigma_{Q}^{2}(0)}.
    \end{aligned}
  \right.
  \label{hi(0)}
\end{equation}

Since it has been shown in \citep{LindaInflation} that the value of $S$ has no influence on the duration of slow-roll inflation, it will be set to zero in the following.

Finally, the initial Hubble parameter can also be expressed as

\begin{equation}
H(0) = - \sqrt{ \sigma_{Q}^{2}(0)+\dfrac{\kappa}{3}\rho(0)-\lambda^{2}\gamma^{2}\left(\dfrac{3}{2}\sigma_{Q}^{2}(0)+\dfrac{\kappa}{3}\rho(0)\right)^{2} }.
\label{H(0)friedmann}
\end{equation}

Equations (\ref{hi(0)}) and (\ref{H(0)friedmann}) define the initial conditions for the background dynamics.

\subsubsection{Simulations}

\begin{figure}[!h]
\begin{center}
\includegraphics[scale=0.35]{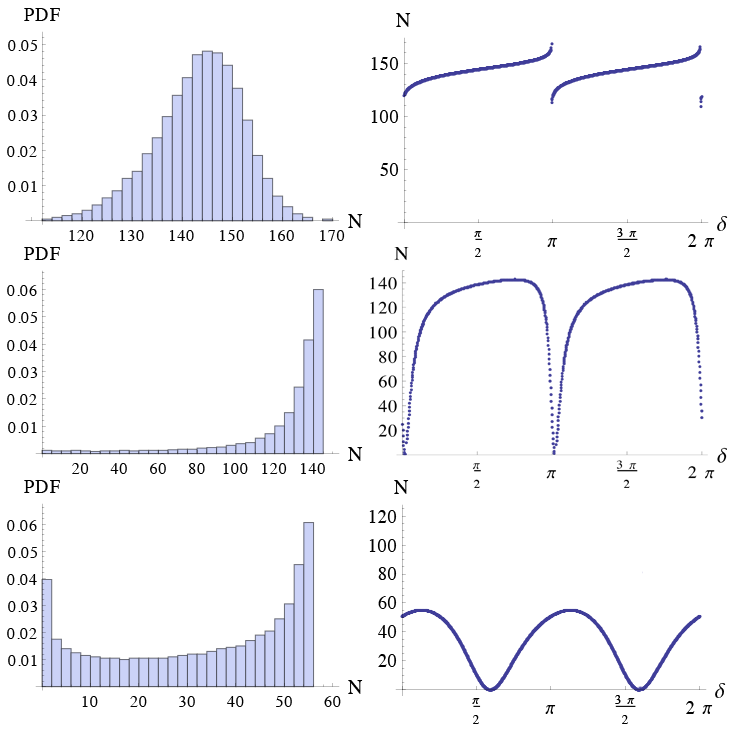}
\caption {Quadratic inflaton potential with initial conditions set before the bounce. \underline{Left column}: Probability distribution functions for the number of \textit{e}-folds of inflation. \underline{Right column}: Number of \textit{e}-folds of inflation as a function of the initial phase of the inflaton field. \underline{Upper panels}: Isotropic universe $f=0$. \underline{Middle panels}: Anisotropic universe, $f=10^{-4}$. \underline{Lower panels}: Anisotropic universe, $f=10^{-2}$.} 
\label{Quadratic CI contraction}
\end{center}
\end{figure}

\begin{figure}[!h]
\begin{center}
\includegraphics[scale=0.35]{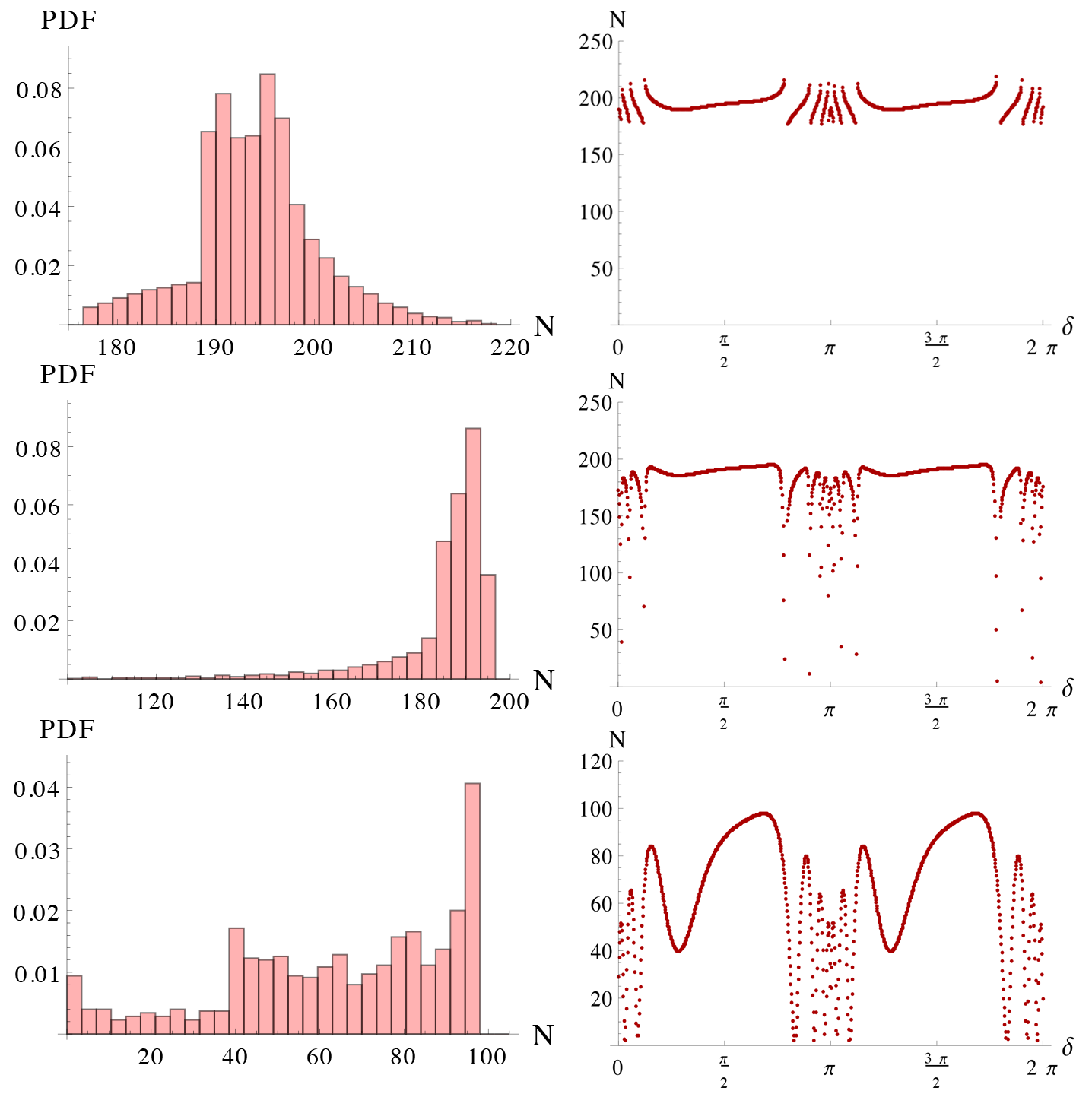}
\caption {Linear inflaton potential with initial conditions set before the bounce. \underline{Left column}: Probability distribution functions for the number of \textit{e}-folds of inflation. \underline{Right column}: Number of \textit{e}-folds of inflation as a function of the initial phase of the inflaton field. \underline{Upper panels}: Isotropic universe $f=0$. \underline{Middle panels}: Anisotropic universe, $f=10^{-4}$. \underline{Lower panels}: Anisotropic universe, $f=10^{-2}$.} 
\label{Monomial CI contraction}
\end{center}
\end{figure}

\begin{figure}[!h]
\begin{center}
\includegraphics[scale=0.35]{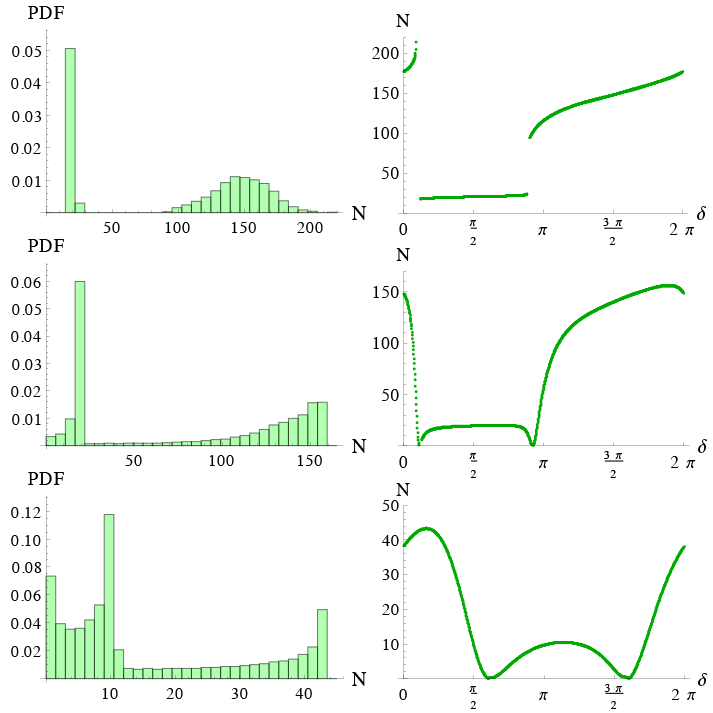}
\caption {String theory inflaton potential with initial conditions set before the bounce. \underline{Left column}: Probability distribution functions for the number of \textit{e}-folds of inflation. \underline{Right column}: Number of \textit{e}-folds of inflation as a function of the initial phase of the inflaton field. \underline{Upper panels}: Isotropic universe $f=0$. \underline{Middle panels}: Anisotropic universe, $f=10^{-4}$. \underline{Lower panels}: Anisotropic universe, $f=10^{-2}$.} 
\label{Stringy CI contraction}
\end{center}
\end{figure}

\begin{figure}[!h]
\begin{center}
\includegraphics[scale=0.35]{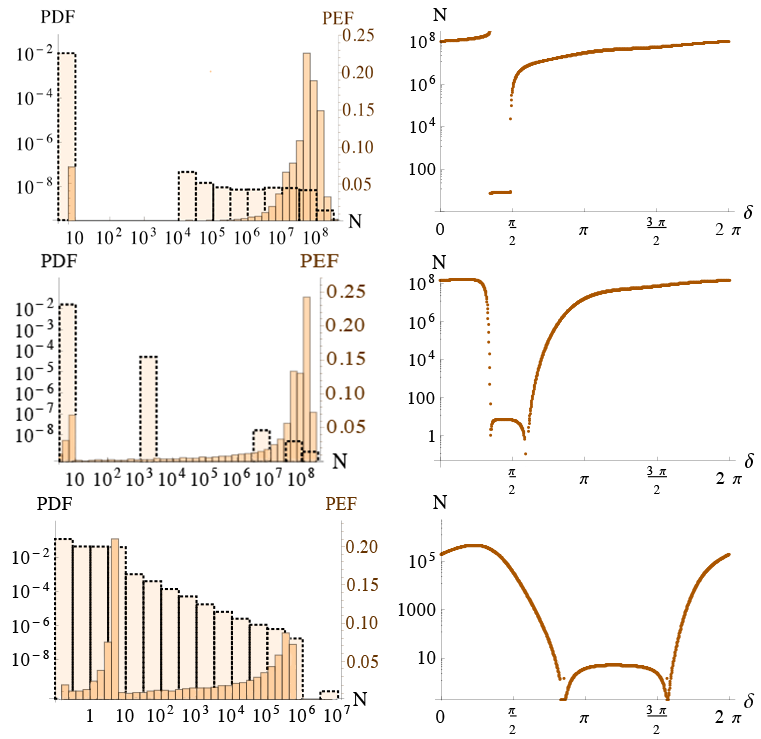}
\caption {Starobinsky potential with initial conditions set before the bounce. \underline{Left column}: Probability distribution functions (and PEFs when useful) for the number of \textit{e}-folds of inflation. \underline{Right column}: Number of \textit{e}-folds of inflation as a function of the initial phase of the inflaton field. \underline{Upper panels}: Isotropic universe $f=0$. \underline{Middle panels}: Anisotropic universe, $f=10^{-4}$. \underline{Lower panels}: Anisotropic universe, $f=10^{-2}$.} 
\label{Starobinsky CI contraction}
\end{center}
\end{figure}

The histograms in the first columns of Figs. \ref{Quadratic CI contraction}, \ref{Monomial CI contraction}, \ref{Stringy CI contraction} and \ref{Starobinsky CI contraction} are estimators of the PDFs of the duration of slow-roll inflation, with respect to the measure $dN$, and for different values of the initial rate of anisotropies: $f=0$, $f=10^{-4}$ and $f=10^{-2}$. They tend toward the real PDFs in the limits $\Delta \delta \to 0$ and $\Delta N \to 0$. The second columns of those figures represent the duration of inflation as a function of the initial phase of the inflaton field for a given value of $f$. This investigation has already been performed for a quadratic potential \citep{LindaInflation}, and it was shown that, as anisotropies grow up, the mean value of the PDF for the number of \textit{e}-folds decreases. We recover this result in Fig. \ref{Quadratic CI contraction}. For high amounts of shear, the distribution becomes bimodal, one side corresponding to ``energy-dominated" bounces and the other one to ``shear-dominated" bounces.\\

An important comment is here in order. When considering models leading to very high numbers of \textit{e}-folds, a logarithmic scale is useful for a better visualization of the full dynamics. In this case, however, the usual PDF normalization fails to capture the most important feature. The standard normalization is indeed such that the sum of the contents of each bin multiplied by its width is equal to 1. With this choice, the contents of the last bins -- when using a log scale -- will be very suppressed in the plot just because the width is large, thus giving the wrong feeling that a high number of \textit{e}-folds is improbable. For this reason, when using a logarithmic scale, we superimpose on each plot the PDF and what we call the probability estimator function (PEF). This estimator uses a normalization such that the sum of the contents of the bins is unitary. Although not strictly a PDF this estimator is more intuitive and allows the reader to immediately see what is the most probable number of \textit{e}-folds. We recommend to base the conclusions on the PEF rather than on the PDF when both are given. When using a linear scales, both estimators coincide (with just a different \textit{y} scale). To avoid making the text too heavy we use the term PDF as a generic one in the following. However, when a log scale is used on the plot, the trend which is mentioned will appear more clearly on the PEFs. When a PDF is represented alone on a plot it is always a solid line; however when it is superimposed with a PEF, the PDF is then represented as a dotted line to emphasize the clearer interpretation of the PEF.

It can be seen in Figs. \ref{Monomial CI contraction}, \ref{Stringy CI contraction} and \ref{Starobinsky CI contraction} that this trend also appears for the other potentials. This, however, is not surprising when considering models with anisotropic shear: the three scale factors associated to the three spatial directions will not reach their minimum value at the same time during the contraction phase. Thus, the maximum amount of energy density available for the scalar field during the bouncing phase will be lower than in the isotropic case. The inflaton field will not be pushed along its potential as far as in the isotropic case, leading to a shorter phase of slow-roll inflation. The major effect of anisotropies is therefore not a modification in the dynamical equations of the Universe \footnote{Since anisotropies scale as $a^{-6}$, the dynamics is almost always equivalent to the isotropic LQC one.} but a shift in the maximum amount of energy available for the scalar field at the bounce. It is mainly this effect which leads to a smaller number of \textit{e}-folds of slow-roll inflation.\\

As explained above, one of the most important features of LQC relies on the fact that the duration of slow-roll inflation is well constrained when initial conditions are set in the contracting phase. This remains partially true when anisotropies are taken into account, although the relative widths of the PDFs increase and their mean values decrease. It should however be emphasized that this important feature of LQC is actually only true as long as the inflaton potential is sufficiently confining. If one considers for example the Starobinsky potential, as shown in Fig. \ref{Starobinsky CI contraction}, the period of slow-roll inflation lasts much longer compared to the cases with other potentials. This is because of the large ``plateau". We recall here that this potential has initially been introduced for quantum gravity reasons, and, at the phenomenological level, for obtaining a long enough phase of inflation, even when the energy density remains small. However, the LQC dynamics automatically provides highly energetic field configurations at the onset of inflation. The inflaton field is therefore ``pushed" far away on the plateau, leading to a very long phase of slow-roll inflation. The peak of the PDF (without shear) of the number of \textit{e}-folds around 150-200 \textit{e}-folds, which is generic for confining potentials in LQC, is now shifted to very different values around $10^8$.

In addition, the bimodal shape of the PDFs in the cases of the string theory and of the Starobinsky potentials is due to the fact that those two potentials are highly asymmetric, as described in \cite{StringPotential} and \citep{Bonga2016}. The low-N peaks correspond to cases where the scalar field is negative at the beginning of inflation, i.e. in the region where the potential is sharp. On the other hand, the high-N peak corresponds to positive values of the scalar field at the beginning of inflation, i.e. where the potentials have a plateau.\\

It is important to underline that for all the considered potentials, the way the number of \textit{e}-folds varies with respect to the phase $\delta$ is highly nontrivial. This is one of the reasons why exhaustive simulations are necessary. From the phenomenological viewpoint, it is worth stressing that for all potentials but the Starobinsky potential, the predicted number of \textit{e}-folds, especially when anisotropies are taken into account, is not much higher than the minimum value favored by observations (around 70 \textit{e}-folds). We want to stress that this provides an opportunity to make quantum gravity effects potentially observable. If inflation lasts much longer than 70 \textit{e}-folds, physical modes with the size of a Planck length at the bouncing time become larger than the Hubble radius at present times, which would make the detection of possible quantum gravity effects very difficult, if not hopeless. But if inflation was not much longer than 70 \textit{e}-folds, an interesting window opens up on LQC phenomenology.

\subsection{Initial conditions at the bounce}

In this section, we consider the case in which initial conditions are set at the bounce ($t=0$ now refers to the bouncing time). The variable to which a presumably known PDF can be assigned is no longer the initial phase of the scalar field $\delta$, but the initial potential energy parameter $x(0)$. A flat PDF will be assumed for $x(0)$, as in \cite{AS2011} and many historical studies, although it is far less motivated than the flat PDF for the $\delta$-parameter used in the previous section.\\

\begin{figure}[!h]
\begin{center}
\includegraphics[scale=0.31]{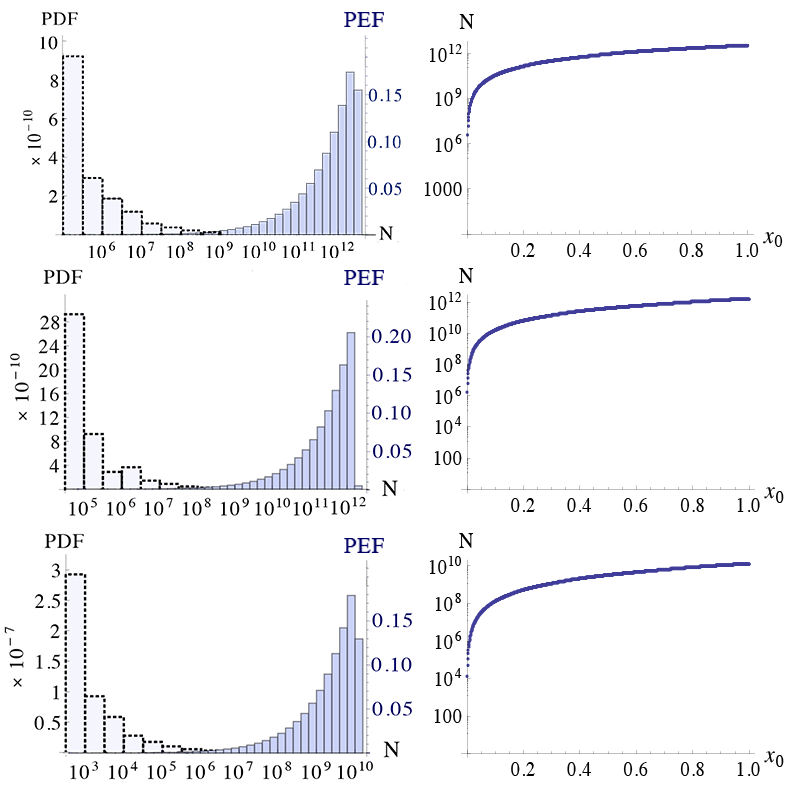}
\caption {Quadratic inflaton potential with initial conditions set at the bounce and positive values of $\Phi(0)$ and $\dot{\Phi}(0)$. \underline{Left column}: Probability distribution functions  (and PEFs) for the number of \textit{e}-folds of inflation. \underline{Right column}: Number of \textit{e}-folds of inflation as a function of $x_{0}$. \underline{Upper panels}: Isotropic universe $f=0$. \underline{Middle panels}: Anisotropic universe, $f=0.57$. \underline{Lower panels}: Anisotropic universe, $f=118$.} 
\label{Quadratic CI bounce}
\end{center}
\end{figure}

\begin{figure}[!h]
\begin{center}
\includegraphics[scale=0.30]{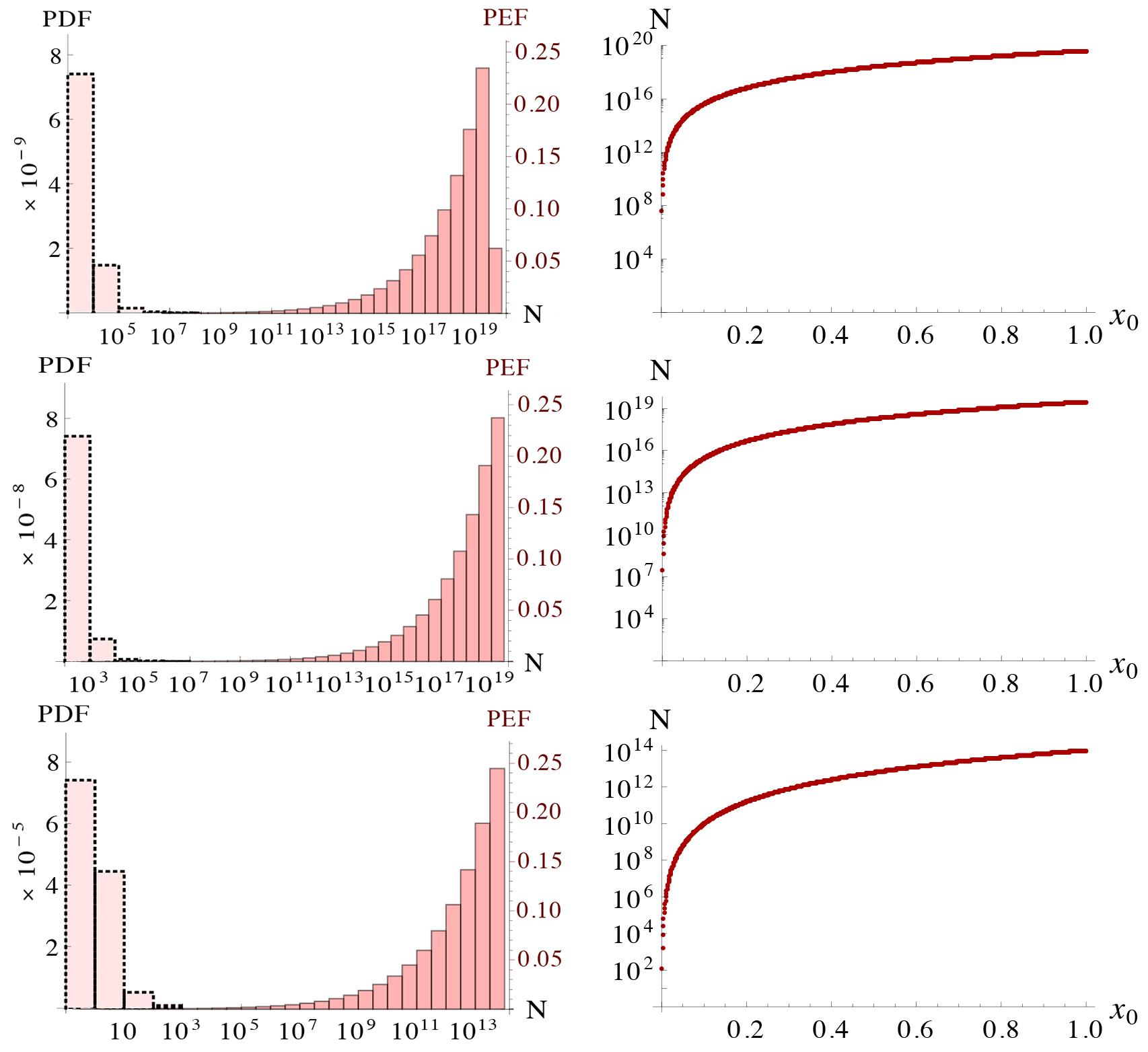}
\caption {Linear inflaton potential with initial conditions set at the bounce and positive values of $\Phi(0)$ and $\dot{\Phi}(0)$. \underline{Left column}: Probability distribution functions (and PEFs) for the number of \textit{e}-folds of inflation. \underline{Right column}: Number of \textit{e}-folds of inflation as a function of $x_{0}$. \underline{Upper panels}: Isotropic universe $f=0$. \underline{Middle panels}: Anisotropic universe, $f=8.73 \times 10^{-2}$. \underline{Lower panels}: Anisotropic universe, $f=288$.} 
\label{Monomial CI bounce}
\end{center}
\end{figure}

The initial shear is still introduced as a fraction of the initial energy density, $\sigma_{Q}^{2}(0) = f ~ \kappa / 3 ~ \rho(0) $, in order to be able to properly compare the effects of anisotropies with what happened in the previous case, where initial conditions were set before the bounce. The initial value of $f$ is obtained by averaging its values over all phases at the bounce in the case of initial conditions set in the remote past.   

The value of the initial energy density can easily be calculated:

\begin{eqnarray}
& & H(0)=0 \\ \nonumber 
& \Leftrightarrow & \sigma_{Q}^{2}(0)+\dfrac{\kappa}{3}\rho(0)-\lambda^{2}\gamma^{2}\left(\dfrac{3}{2}\sigma_{Q}^{2}(0)+\dfrac{\kappa}{3}\rho(0)\right)^{2} =0  \\ \nonumber
& \Leftrightarrow & \rho(0) = 3 \dfrac{f+1}{\kappa \lambda^{2} \gamma^{2}} \dfrac{1}{\left( 1 + \dfrac{3}{2}f \right)^{2}}.
\end{eqnarray}

To obtain the initial conditions for the $h_{i}$-coefficients, we fix one of them to $h_{i}(0)= n \pi/2, ~ n \in \mathbb{N},~ i=1,2,3 $ and the two others ($h_{j}(0)$ and $h_{k}(0)$, $j,k = 1,2,3$ , $i \neq j \neq k$)  are then fixed by the following constraints,

\begin{equation}
\sin(h_{1}+h_{2})+\sin(h_{1}+h_{3})+\sin(h_{2}+h_{3}) = 0,
\end{equation}

and

\begin{equation}
\cos(h_{1}-h_{2})+\cos(h_{1}-h_{3})+\cos(h_{2}-h_{3}) = 3 - 9 f \dfrac{f+1}{(1+3f/2)^{2}} ~,
\end{equation}

obtained from Eqs. (\ref{H(hi)}) and (\ref{SigmaQ(hi)}). One of the $h_{i}(0)$'s must be a multiple of $\pi$ otherwise solutions to this system are nonreal.\\

Figures \ref{Quadratic CI bounce} and \ref{Monomial CI bounce} display the results of the simulations obtained for the quadratic and the linear potentials, for positive values of $\Phi(0)$ and $\dot{\Phi}(0)$. It is clear  that anisotropies have  no significant effects on the shapes of the probability distribution functions for $N$. However, the mean value of $N$ decreases when $f$ increases, similar to the behavior of $N$ when initial conditions are set in the prebouncing phase. This is not surprising since the major effect of the shear is a decrease in the energy density available for the scalar field at the bounce. It should also be underlined that $N$ increases significantly when $x(0)$ grows up. Those  large values of $N$ were nearly never reached in the previous scenario, when initial conditions are set before the bounce, because a very high level of fine-tuning of the initial phase $\delta$ would have been required to generate a nontiny value of $x(0)$. Obviously, the duration of inflation is less constrained when initial conditions are set at the bounce with a flat PDF on $x(0)$. The total number of \textit{e}-folds is much higher than $N^{\star}\sim 60-70$ which would correspond to visible inflation.\\

It should be underlined that a flat PDF for $x(0)$ may not be relevant when setting initial conditions in the case of nonsymmetric potentials, such as the string theory potential or the Starobinsky potential. For those potentials, a given value of  $V(\Phi)$ corresponds to two different values of $|\Phi|$, and consequently to two different evolutions of the scalar field. 

\begin{figure}[!h]
\begin{center}
\includegraphics[scale=0.31]{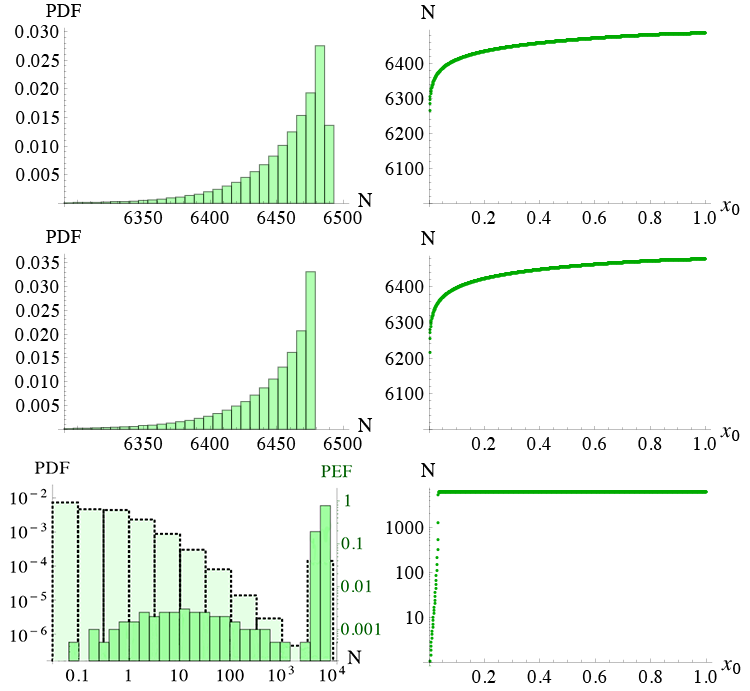}
\caption {String-theory inflaton potential with initial conditions set at the bounce and positive values of $\Phi(0)$ and $\dot{\Phi}(0)$. \underline{Left column}: Probability distribution functions (and PEFs when useful)  for the number of \textit{e}-folds of inflation. \underline{Right column}: Number of \textit{e}-folds of inflation as a function of $x_{0}$. \underline{Upper panels}: Isotropic universe $f=0$. \underline{Middle panels}: Anisotropic universe, $f=0.31$. \underline{Lower panels}: Anisotropic universe, $f=355$.} 
\label{Stringy CI bounce}
\end{center}
\end{figure}

The case of the string theory potential is presented in Fig. \ref{Stringy CI bounce}. Positive values of $\Phi(0)$ and $\dot{\Phi}(0)$ were chosen in order to probe the right part of the potential, and in order to be comparable with the two previous potentials. The first two lines show that in the cases $f=0$ and $f=0.31$, the duration of slow-roll inflation does not vary a lot with $x(0)$. This behavior is due to the fact that for nearly all the displayed values of $x(0)$, the value of the potential energy at the beginning of the slow-roll phase is higher than the plateau. 
On the third line, however, the amount of shear becomes high enough so that, at low $x(0)$, the potential energy becomes lower than the plateau. This implies much shorter durations of inflation.

\begin{figure}[!h]
\begin{center}
\includegraphics[scale=0.31]{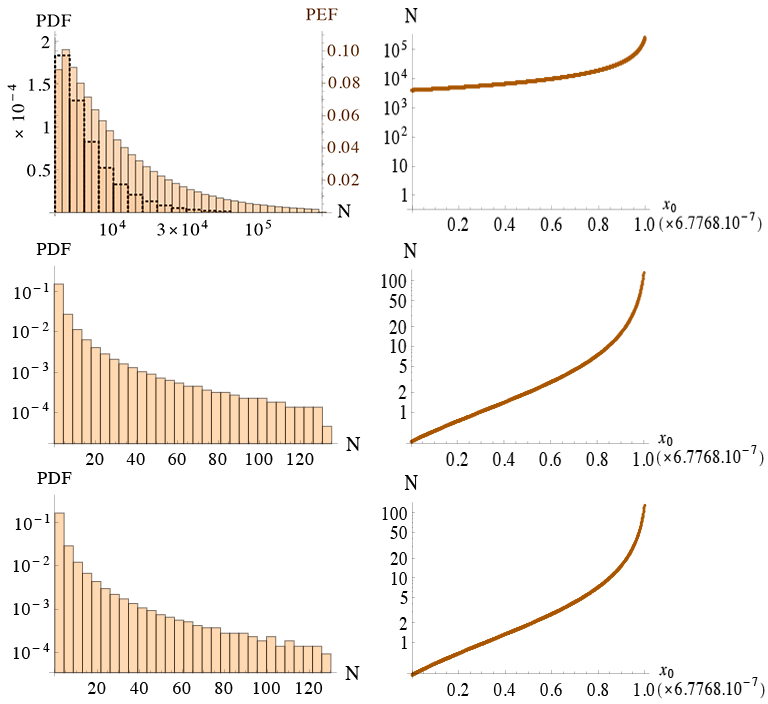}
\caption {Starobinsky inflaton potential with initial conditions set at the bounce and positive values of $\Phi(0)$ and $\dot{\Phi}(0)$. \underline{Left column}: Probability distribution functions (and PEFs when useful) for the number of \textit{e}-folds of inflation. \underline{Right column}: Number of \textit{e}-folds of inflation as a function of $x_{0}$. \underline{Upper panels}: Isotropic universe $f=0$. \underline{Middle panels}: Anisotropic universe, $f=3.87$. \underline{Lower panels}: Anisotropic universe, $f=28.0$.} 
\label{Starobinsky CI bounce}
\end{center}
\end{figure}

The case of the Starobinsky potential is slightly more complicated. The initial value of the inflaton field $\Phi(0)$  can be expressed as a function of $x(0)$:

\begin{equation}
\Phi(0) = - \sqrt{\dfrac{3}{2 \kappa}} \log \left( 1\mp \sqrt{\dfrac{4 \kappa \rho_{c}}{3 m^{2}}} x(0) \right),
\end{equation}

where the ``minus" solution in the logarithm corresponds to positive values of $\Phi(0)$ whereas the ``plus" solution corresponds to negative ones. If we consider positive values of $\Phi(0)$, a specific value of $x(0)$ appears:

\begin{equation}
x_{c}(0) = \sqrt{\dfrac{3m^{2}}{4 \kappa \rho_{c}}} = 6.77\times 10^{-7}.
\end{equation}

It corresponds to the value of $x(0)$ for which the potential energy is equal to the value of the plateau of the potential: $x_{c}(0)=\sqrt{\dfrac{V_{\text{plate}}}{\rho_{c}}}$.

We distinguish two cases:

\begin{itemize}
\item \underline{$x(0) < x_{c}(0)$}: For those values of $x(0)$, the initial potential energy density at the bounce is lower than the plateau. As mentioned previously, a single value of  $V$ corresponds to two different values of $\Phi$, one of them being positive and the other one negative.
\item \underline{$x(0) > x_{c}(0)$}: These values of  $x(0)$ correspond to potential energy densities which are higher than the plateau. For a given value of $V$, there is now only one negative value of $\Phi$.
\end{itemize}

Since $x_{c}(0) \ll 1$, if one wants to vary $x(0)$  between 0 and 1, and probe the plateau part of the potential, it is necessary to take negative values of $\Phi(0)$, together with positive values of $\dot{\Phi}(0)$. It remains possible to probe the plateau with positive values of $\Phi(0)$ and $\dot{\Phi}(0)$ if  $x(0) \in [0,x_{c}(0)]$. It should be noticed that, if initial conditions are set in the contracting phase, positive values of the field at the bounce are highly favored in the isotropic case, and remain favored in the presence of anisotropic shear, as shown in Fig. \ref{Starobinsky CI contraction} \footnote{In most cases, the field has the same sign at the bounce and at the beginning of the slow-roll phase.}. We therefore choose to show some results associated with $\Phi(0)$ and $\dot{\Phi}(0)$ with  $x(0) \in [0,x_{c}(0)]$ in Fig. \ref{Starobinsky CI bounce}. It can be seen that without shear the inflaton field is pushed far away on the plateau, leading to large numbers of \textit{e}-folds. However, when the initial shear is nonvanishing, the energy density which remains available for the scalar field is smaller, such that the field cannot reach the plateau anymore. This leads to a shorter slow-roll phase. It is difficult to probe the plateau with a flat PDF for $x(0)$  if anisotropies are taken into account. Since the cosmological interest of the Starobinsky potential is mostly associated with the plateau, this means that setting initial conditions at the bounce, at least in the presented way, is not very relevant in this case.\\
  
From the viewpoint of phenomenology, it is important to notice that the predicted number of \textit{e}-folds, if initial conditions are believed to be set at the bounce, is generically very high. Unless a huge amount of fine-tuning is applied, the observation of possible quantum gravity effects in the CMB is virtually impossible. Only in the case of a strongly shear-dominated bounce does the number of \textit{e}-folds become close to the observational bound. \\

The usually much smaller number of \textit{e}-folds of inflation when initial conditions are set in the classical prebounce phase, can be understood as follows: setting initial conditions, {\it i.e.} fixing the initial phase of the inflaton field, when the energy density is very small leads -- for almost all values of the phase -- to solutions without deflation. This has already been (implicitly) shown in the frame of standard cosmology by Gibbons and Turok \cite{Gibbons:2006pa} and the consequences of these results for LQC were explained in detail in \cite{Bolliet:2017czc}. Solutions to the given set of differential equations without deflation cannot bring the field to high values at the bounce, since the accelerated contraction stops almost immediately. One therefore encounters a kinetic-energy dominated bounce which subsequently leads to a small number of \textit{e}-folds, as shown in Fig. 2 of \cite{Bolliet:2017czc}. On the other hand, varying the value of the field  at the bounce -- hence making potential-energy dominated bounce scenarios likely -- results in very large numbers of e-folds for many solutions. The above arguments were shown for the quadratic potential, but they still hold for the linear potential and the string-theory-inspired potentials. Taking anisotropies into account makes the energy density available for the scalar field  even smaller and subsequently decreases the resulting number of \textit{e}-folds for a particular solution.

\section{Discussion and Conclusion}

\subsection{Discussion}

Let us begin by discussing the issue of the best choice for initial conditions. If the word ``initial" is taken in its literal sense, it is certainly reasonable to set them in the remote past and respect the causal evolution of the system. As shown in \cite{bl}, the evolution across the bounce is not time symmetric.  From the mathematical viewpoint this is however not necessary and some physical arguments are required. It seems to us that assigning a flat PDF to the phase of the field in the remote past of the contracting branch is a better choice than assigning a flat PDF  to the fraction of potential energy at the bounce. The first reason for this is that the vicinity of the bounce is the most ``quantum" period in the history of the Universe. It is therefore the one where the semiclassical approach used here is the most questionable -- backreaction might not be negligible -- and hence the worst one to assign specific values to the dynamical variables. This is precisely the time when the considered system is not under perfect control and obviously not the most natural one to set initial conditions in a safe way. The second reason is that a flat PDF for the fraction of potential energy is a completely arbitrary choice. It has no physical motivation, the PDF could be chosen to be anything else with the same credibility. There is no reason to chose all potential energies with the same probability. Describing the very same system with other variables to which flat PDFs could be assigned would lead to completely different PDFs for the fraction of potential energy and to completely different results for the number of expected \textit{e}-folds. This is to be contrasted with the flat PDF assigned to the phase of the oscillations. In this case, the phase has a clear physical meaning and is a random variable with a known PDF during an oscillatory process. One could discuss the details of the PDF but the rough shape is known just because the field is an oscillator. It could be argued that the fraction of potential energy is also known and this is true, but not at the bounce time where the dynamics is modified with respect to the trivial nearly oscillatory process. The third reason is that a flat PDF assigned to the phase is preserved over time. This is very important and means that this choice is consistent in the sense that it does not depend on the chosen hypersurface at which initial conditions are set. Obviously, a flat PDF for the fraction of potential energy is not time preserved and there is no reason for the bounce to be the precise time when assigning a flat PDF to this variable.

Knowing the PDF for any dynamical variable describing the system allows one to know the PDF for the number of \textit{e}-folds. There are two kinds of ``predictive powers" that need to be distinguished at this stage. Let us call ``strong predictive power" the case in which the number of \textit{e}-folds of inflation is (roughly) known and ``weak predictive power" the case in which the PDF for the number of \textit{e}-folds is known. The strong case basically requires that the PDF is not only known (that is, the weak case) but also requires that it is highly peaked.\\

\subsection{Conclusion}

This study is dedicated to the systematic investigation of the duration of inflation in LQC with holonomy corrections. We have addressed the three main unknown points: the way to set initial conditions, the amount of shear and the shape of the inflaton potential. The conclusions of this study are the following: (i) As far as the capability of the model to predict the distribution of the number of \textit{e}-folds is concerned, it is, in our opinion, more appealing to set initial conditions in the remote past of the classical contracting branch of the Universe. In this case, a flat PDF can easily be associated to the $\delta$-parameter for all the potentials. (ii) Furthermore, in this case, the duration of inflation is indeed severely constrained, and most interestingly to values which are not much higher than the minimum value required by observations (but only for ``confining" potentials). (iii) When anisotropies are taken into account the PDF of the number of e-folds is widened and its mean value decreases confirming the strong predictive power of LQC for a massive scalar field. (iv) For potentials with a plateau such that the favored value of the amount of potential energy at the beginning of the slow-roll phase is larger than the height of the plateau, the predicted number of \textit{e}-folds can become very large and the predictive power is only weak. (v) When the potential is asymmetric, the PDF can become bimodal. (vi) When initial conditions are set at the bounce, even the weak predictive power of LQC is basically lost as everything is then determined by the arbitrary choice of the variable to which a known PDF is assigned.\\

In summary, if the shape of the inflaton potential can be experimentally determined (this is already partially the case) and if, following the logics of causality, the initial conditions are set in the remote past, there is an obviously interesting predictive power of LGC for the duration of inflation. This predictive power is strong if the potential is confining and weak if the potential has a plateaulike shape. It is not so because of the specific quantum dynamics but because of the existence of a preferred amount of potential energy at the onset of inflation which is naturally selected by the semiclassical trajectory. The most difficult point to address remains the one of anisotropies as no simple physical argument allows one to choose a preferred amount of shear. If the potential is confining enough this is however not necessarily a problem as the predicted number of \textit{e}-folds is then restricted to a quite small interval (bounded from above by the model in the isotropic case and from below by observations as $N>N^{\star}\approx 70$) which happens to be the most interesting one for phenomenology.

\section*{Acknowledgments}

We thank B. Bolliet for helpful discussions. K.M is supported by a grant from the CFM foundation. S.S. is supported by grants from the Heinrich-B\"{o}ll-Stiftung e.V. and the Studienstiftung des deutschen Volkes e.V.

\cleardoublepage

\bibliography{refs}
\end{document}